\documentclass{article}
\usepackage{spconf,amsmath,graphicx,hyperref}
\usepackage{enumitem}
\usepackage{booktabs}
\usepackage{multirow} 
\usepackage{adjustbox}
\usepackage{xcolor}    
\usepackage{pifont}     
\usepackage{makecell}   
\usepackage{siunitx}
\usepackage{float}

\hypersetup{
  colorlinks=true,    
  citecolor=green,    
  filecolor=magenta   
}

\title{VStyle: A Benchmark for Voice Style Adaptation\\ with Spoken Instructions}

\name{%
  \begin{tabular}{c}  
    Jun Zhan$^{1,2,*}$\thanks{* Equal contribution.}, 
    Mingyang Han$^{2,*}$, 
    Yuxuan Xie$^{1,*}$, 
    Chen Wang$^{2}$, Dong Zhang$^{1}$, \\
    Kexin Huang$^{1}$, Haoxiang Shi$^{2}$, DongXiao Wang$^{2}$, Tengtao Song$^{2}$, \\
    Qinyuan Cheng$^{1}$, Shimin Li$^{1}$, 
    Jun Song$^{2,\dagger}$\thanks{$\dagger$ Corresponding authors.}, Xipeng Qiu$^{1,\dagger}$, 
    Bo Zheng$^{2}$
  \end{tabular}}

\address{%
  \begin{tabular}{c}
    $^{1}$Fudan University \quad $^{2}$Alibaba Group \\
    \small \texttt{jzhan24@m.fudan.edu.cn, xpqiu@fudan.edu.cn} \\
    \small \texttt{\{hanmingyang.hmy, jsong.sj\}@alibaba-inc.com}
  \end{tabular}
}

\begin{document}
%
\maketitle

\begin{abstract}

Spoken language models (SLMs) have emerged as a unified paradigm for speech understanding and generation, enabling natural human–machine interaction. However, while most progress has focused on semantic accuracy and instruction following, the ability of SLMs to adapt their speaking style based on spoken instructions has received limited attention. We introduce Voice Style Adaptation (VSA), a new task that examines whether SLMs can modify their speaking style—such as timbre, prosody, or persona—following natural language spoken commands.
To study this task, we present VStyle, a bilingual (Chinese \& English) benchmark covering four categories of speech generation: acoustic attributes, natural language instruction, role play, and implicit empathy. We also introduce the Large Audio Language Model as a Judge (LALM-as-a-Judge) framework, which progressively evaluates outputs along textual faithfulness, style adherence, and naturalness, ensuring reproducible and objective assessment.
Experiments on commercial systems and open-source SLMs demonstrate that current models face clear limitations in controllable style adaptation, highlighting both the novelty and challenge of this task. By releasing VStyle and its evaluation toolkit, we aim to provide the community with a foundation for advancing human-centered spoken interaction. The dataset and code are publicly available at \href{https://junzhan2000.github.io/VStyle.github.io/}{project's homepage}.

\end{abstract}


\section{Introduction}
\label{sec:intro}

Spoken language models (SLMs)\cite{arora2025landscape, cui2024recent, zhang2023speechgpt} have recently gained wide attention. Compared with traditional cascaded pipelines, they offer more natural and realistic interactions. However, most research focuses on \textbf{what the model says} (semantic accuracy) rather than \textbf{how the model says it} (expressiveness). Non-verbal cues---such as speaker identity, emotion, and paralinguistic signals---are crucial to natural dialog and user experience, but comprehensive frameworks for evaluating these expressive dimensions remain lacking.

Conventional TTS metrics\cite{tan2021survey} like Word Error Rate (WER) and Speaker Similarity (SIM) fail to adequately capture the diverse requirements of dialogue systems. Existing benchmarks have limitations: Some\cite{hassid2023textually, chen2024voicebench} consider only content accuracy; VocalBench\cite{liu2025vocalbench} targets conversational style, not speech style; S2S-Arena\cite{jiang2025s2s} relies on costly, unscalable human evaluations; WavReward\cite{ji2025wavreward} lacks coverage of realistic dialog scenarios; and Kimi-Audio\cite{ding2025kimi}, Step-Audio\cite{huang2025step,wu2025step} provide only small-scale test sets with limited reproducibility.

To address these gaps, we first formalize the task of \textbf{Voice Style Adaptation (VSA)}: determining whether an SLM can modify its speaking style in response to spoken instructions. To study this task, we introduce \textbf{VStyle}, a bilingual (Chinese/English) benchmark covering four categories of speech generation: acoustic attributes, natural-language instructions, role-play, and implicit empathy, comprising 1,523 prompts designed around realistic interaction needs.

A key challenge is reliable quantitative evaluation, since human assessment is costly and variable. We therefore introduce the \textbf{LALM-as-a-Judge} framework\cite{gu2024survey, chiang2025audio, manku2025emergenttts, huang2025instructttseval}, which leverages Large Audio Language Models to progressively assess outputs across three dimensions: content faithfulness, style consistency, and overall naturalness. This enables a scalable and reproducible automatic evaluation pipeline.

Applying VStyle to commercial and open-source SLMs, we show it effectively distinguishes voice style adaptation performance and reveals a significant gap across systems. We release the dataset and toolkit to support progress toward more expressive, controllable, and human-centered spoken interaction.

\begin{figure*}[h]
    \centering
    \includegraphics[width=0.95\textwidth]{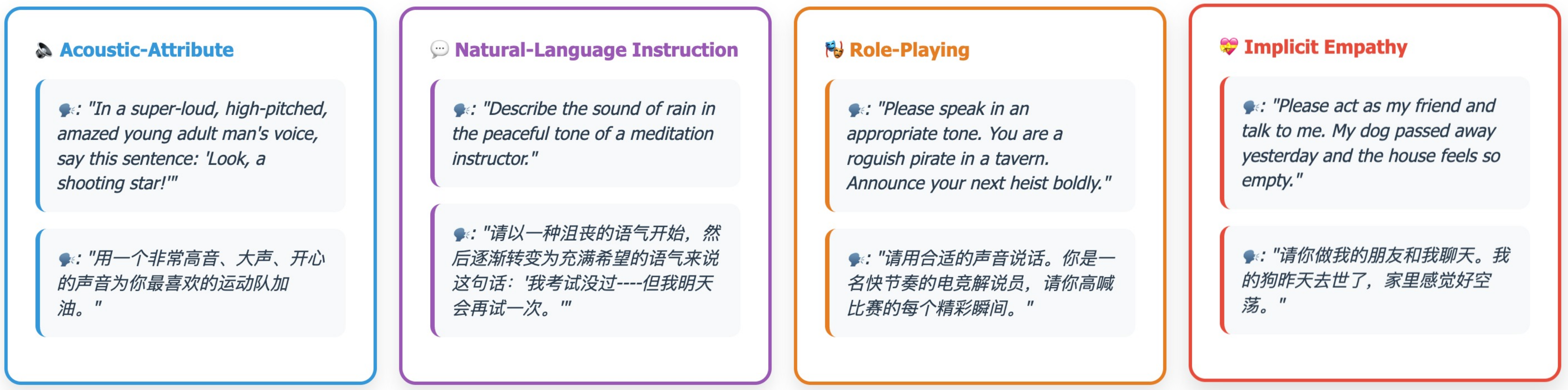} 
    \caption{Instruction examples from the VStyle dataset across four categories: Acoustic Attributes, Natural Language Instruction, Role-Playing, and Implicit Empathy.}
    \label{fig:main} 
\end{figure*}

\section{VStyle}

\label{sec:VStyle}

\subsection{Overview}
VStyle is a bilingual benchmark for evaluating voice style adaptation (VSA) in spoken language models. Each instance provides a spoken instruction that may specify content, assign a task, or set up an interaction, but also the desired speaking style, expressed explicitly or implicitly. The model must generate a spoken response that aligns with both intent and style. To reflect realistic interaction needs, VStyle covers four categories: \textit{acoustic attributes, natural language instruction, role-play, and implicit empathy}, with examples shown in Figure~\ref{fig:main}.

\textbf{Acoustic Attributes.}
In this category, the instruction explicitly constrains one or more acoustic attributes of the generated speech, including age, gender, speaking rate, pitch, loudness, and emotion. Each attribute is defined within a finite closed set, enabling direct evaluation of a model’s capacity for fine-grained yet essential control over speech acoustics.

\textbf{Natural-Language Instructions.} 
This category uses open-ended natural language instructions to guide speaking style generation. It includes three subtypes: \textit{emotion}, referring to unconstrained descriptions of affective states; \textit{style}, allowing free-form specification of global speech style; and \textit{variation}, which entails temporal variation between emotions and styles within a single utterance, thereby showcasing a model’s fine-grained controllability.

\textbf{Role-Play.} 
Role-play tasks are categorized into two subtypes: \textit{scene-based} and \textit{character-based}. The former requires the model to assume a role within a given scenario, while the latter involves imitating personas characterized by distinctive vocal traits. Success in both depends on the model’s ability to infer and produce appropriate timbre, emotion, and speaking style from contextual cues.

\textbf{Implicit Empathy.} 
Emotional companionship is a key application of conversational speech systems. In this category, instructions do not specify a speaking style; instead, they prompt the model to interact as a friend while conveying a strong emotion. The model must infer the user’s affective state and deliver a supportive response that integrates both linguistic content and prosodic expression. Four representative affective contexts are considered: \textit{anger}, \textit{anxiety and fear}, \textit{sadness and disappointment}, and \textit{joy and excitement}.

\begin{figure*}[h]
    \centering
    \includegraphics[width=0.95\textwidth]{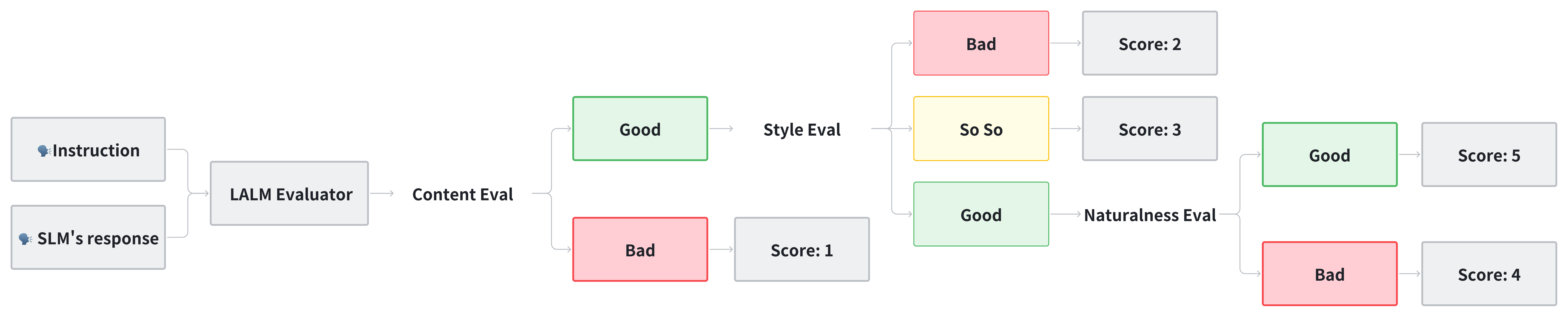} 
    \caption{Evaluation framework using Large Audio-Language Models (LALMs) as a judge. Given a instruction and a generated voice answer, the LALM evaluator conducts a hierarchical assessment: (1) \emph{Content evaluation}, where failure leads to a score of 1; (2) \emph{Style evaluation}, assigning a score of 2 for non-compliance, 3 for partial compliance, and progressing if satisfied; and (3) \emph{Naturalness evaluation}, yielding a score of 4 for unnatural and 5 for highly natural speech. This staged process underlies the 5-point MOS scoring scheme.}
    \label{fig:score} 
\end{figure*}

\subsection{Data Construction}
We constructed the instruction dataset using a hybrid human–LLM approach. Seed instructions were first manually designed, then expanded and iteratively refined with LLMs. To minimize ambiguity in spoken language, we adopted fixed patterns such as *“Please say this sentence”* (content specification), *“Please speak in an appropriate tone”* (role-play), and *“Please act as my friend and talk to me”* (empathy).  

After preparing the text corpus, we synthesized corresponding audio instructions with commercial voice cloning models. For the first three categories that do not require distinct emotional expression, we selected audio prompts from Seed-TTS with DNSMOS scores above 3.2 and generated diverse audio instructions through a commercial voice cloning system. For empathy-related data, Gemini-TTS was employed, leveraging its instruction control capabilities to produce speech aligned with the intended emotional states. Following filtering of erroneous and ambiguous samples, the final dataset comprised 1,523 bilingual (Chinese–English) speech instructions.  

\subsection{Large Audio-Language Model as Judge}
Automatic and quantitative evaluation of speech generation quality remains a fundamental challenge. Recent studies\cite{chiang2025audio, manku2025emergenttts, huang2025instructttseval} indicate that Large Audio-Language Models exhibit strong capabilities in audio assessment, which we adapt for spoken dialogue evaluation.

We structure the evaluation along three hierarchical dimensions: textual adherence, stylistic adherence, and overall naturalness.A 5-point Mean Opinion Score (MOS) scale is employed, with the evaluation procedure illustrated in Figure~\ref{fig:score}. If the generated speech fails to satisfy textual adherence, it is directly assigned a score of 1. Otherwise, the process proceeds to stylistic adherence: a score of 2 is given for complete non-compliance, 3 for partial but imperfect compliance, and full compliance advances the evaluation to naturalness. When the generated speech demonstrates a high level of naturalness, it is awarded the maximum score of 5.

This progressive framework reflects the incremental requirements of speech models: first correctness of content, then stylistic appropriateness, and finally natural, human‑like speech capable of sustaining realistic interaction. Leveraging LALMs’ perceptual and reasoning abilities, our chain‑of‑thought evaluation achieves consistent and stable scoring. Final benchmark scores are obtained by averaging across categories with equal weights.

\section{Experiment}
\label{sec:exp}

\begin{table*}[t]
\caption{Evaluation Results on VStyle. Bold indicates the best score in each category for each language. Abbreviations: Gend. = Gender, Emot. = Emotion, Vol. = Volume, Comp. = Composite, Vari. = Variation, Scen. = Scenario, Char. = Character, Sad. = Sadness/Disappointment, Anx. = Anxiety/Fear, Joy = Joy/Excitement.}
\label{tab:main-results}
\centering
\small
\setlength{\tabcolsep}{3pt}
\renewcommand{\arraystretch}{1.15}
\resizebox{\textwidth}{!}{%
\begin{tabular}{l|c|c|ccccccc|ccc|cc|cccc}
\toprule
\multirow{2}{*}{Model} & \multirow{2}{*}{Lang} & \multirow{2}{*}{Overall} 
& \multicolumn{7}{c|}{Acoustic Attributes} 
& \multicolumn{3}{c|}{Instruction} 
& \multicolumn{2}{c|}{Role-Play} 
& \multicolumn{4}{c}{Empathy} \\
\cmidrule(lr){4-10} \cmidrule(lr){11-13} \cmidrule(lr){14-15} \cmidrule(lr){16-19}
& & & Age & Speed & Gend. & Emot. & Pitch & Vol. & Comp. & Emot. & Style & Vari. & Scen. & Char. & Anger & Sad. & Anx. & Joy \\
\midrule
\multirow{2}{*}{Baichuan-Audio} 
& en & 2.50 & 2.71 & 2.20 & \textbf{3.83} & 2.58 & 2.05 & 2.05 & 2.55 & 2.23 & 2.21 & 1.88 & 2.08 & 2.33 & 2.41 & 3.43 & 2.74 & 3.91 \\
& zh & 2.25 & 2.67 & 2.45 & 3.08 & 2.29 & 2.00 & 2.80 & 2.58 & 1.71 & 1.72 & 1.69 & 2.29 & 1.95 & 2.11 & 2.55 & 2.20 & 3.51 \\
\midrule
\multirow{2}{*}{Step-Audio} 
& en & 2.77 & 2.71 & 2.40 & 2.38 & 2.46 & 2.20 & 3.15 & 2.28 & 2.59 & 2.89 & 2.56 & 1.65 & 2.11 & 3.95 & 4.37 & 3.87 & 4.29 \\
& zh & 2.92 & 2.88 & 2.60 & 2.58 & 3.12 & 2.80 & 2.75 & 2.57 & 2.39 & 2.32 & 2.07 & 2.93 & 2.80 & 3.59 & 4.52 & 3.20 & 4.26 \\
\midrule
\multirow{2}{*}{Qwen2.5-Omni} 
& en & 2.46 & 2.58 & 2.25 & 1.92 & 3.04 & 1.95 & 2.05 & 2.30 & 2.67 & 2.87 & 2.36 & 1.80 & 1.68 & 2.95 & 2.73 & 3.55 & 3.43 \\
& zh & 2.97 & 3.21 & 2.45 & 3.12 & 2.62 & 2.35 & 2.55 & 2.30 & 2.53 & 2.38 & 2.07 & 2.54 & 2.24 & 4.64 & 4.28 & 4.77 & \textbf{4.91} \\
\midrule
\multirow{2}{*}{Kimi-Audio} 
& en & 2.54 & 2.79 & 2.45 & 2.54 & 3.04 & 1.55 & 3.00 & 2.33 & 2.19 & 2.41 & 2.33 & 1.73 & 1.72 & 3.59 & 3.97 & 3.65 & 3.46 \\
& zh & 3.11 & 3.33 & 3.45 & 2.25 & 3.75 & 2.95 & 3.25 & 3.17 & 2.66 & 2.74 & 2.43 & 3.01 & 2.23 & 3.86 & 3.86 & 3.80 & 4.57 \\
\midrule
\multirow{2}{*}{Doubao} 
& en & 3.63 & \textbf{3.75} & 3.55 & 3.46 & 3.38 & 3.25 & 4.05 & 3.13 & 3.52 & 3.67 & 2.90 & 3.27 & 2.56 & 4.89 & 5.00 & 4.81 & 4.94 \\
& zh & \textbf{4.10} & \textbf{3.88} & \textbf{4.35} & 3.25 & \textbf{4.65} & \textbf{4.35} & \textbf{4.70} & \textbf{3.77} & \textbf{3.90} & \textbf{3.96} & 2.88 & \textbf{4.45} & 3.79 & 4.59 & 4.72 & \textbf{4.80} & 4.83 \\
\midrule
\multirow{2}{*}{GPT-4o-Mini} 
& en & 3.88 & 2.83 & \textbf{3.75} & 3.50 & 3.79 & 3.10 & \textbf{4.15} & 3.05 & 3.72 & 4.00 & 3.47 & 3.23 & 3.82 & \textbf{4.98} & \textbf{5.00} & 4.87 & \textbf{5.00} \\
& zh & 3.74 & 3.25 & 3.75 & \textbf{3.50} & 3.75 & 3.30 & 3.70 & 3.32 & 3.46 & 3.47 & 2.98 & 3.48 & 3.84 & 4.30 & 4.52 & 4.73 & 4.69 \\
\midrule
\multirow{2}{*}{GPT-4o} 
& en & \textbf{4.05} & 3.67 & 3.45 & 2.79 & \textbf{4.00} & \textbf{3.60} & 4.10 & \textbf{3.27} & \textbf{3.93} &  \textbf{4.23} & \textbf{4.07} & \textbf{3.89} & \textbf{3.83} & 4.95 & 4.90 & \textbf{5.00} & 4.54 \\
& zh & 3.84 & 3.42 & 3.10 & \textbf{3.50} & 3.83 & 3.35 & 3.90 & 3.22 & 3.37 & 3.51 & \textbf{3.11} & 3.89 & \textbf{3.90} & \textbf{4.75} & \textbf{4.83} & 4.67 & 4.80 \\
\bottomrule
\end{tabular}%
}
\vspace{0.5em}
\end{table*}

\subsection{Experimental Setup}

We evaluate GPT-4o Audio (snapshot: gpt-4o-audio-preview-2025-06-03)~\cite{openai2025audio}, 
GPT-4o-Mini Audio (snapshot: gpt-4o-mini-audio-preview-2024-12-17)~\cite{openai2025audio}, 
and Doubao~\cite{doubao2025} as representative commercial voice-based dialogue systems.
In addition, we include four open-source end-to-end speech–language models noted for their strong speech generation capabilities: 
Step-Audio~\cite{huang2025step}, 
Kimi-Audio~\cite{ding2025kimi}, 
Baichuan-Audio~\cite{li2025baichuan}, 
and Qwen-2.5 Omni~\cite{xu2025qwen2}.

For models supporting multiple fixed speakers, we randomly assigned one speaker per dialogue session. For Baichuan‑Audio, we removed the voice‑prompt to allow more diverse speech generation.
Full‑duplex systems such as Doubao autonomously manage response timing. To ensure outputs were produced only after complete input reception, we avoided inserting long pauses in the speech input.

For evaluation, we employed Gemini‑2.5‑pro, the strongest audio evaluation model validated in prior work\cite{chiang2025audio, manku2025emergenttts}, with inference parameters set to a maximum token length of 4096, a temperature of 1.0, and top‑p of 0.7.

\subsection{Results and Analysis}
\textbf{Overall Results}. 
Table~\ref{tab:main-results} presents the comprehensive evaluation results across all models and evaluation dimensions. Several key findings emerge from our analysis:

\textbf{Significant performance gap between open-source and commercial models}. Commercial models clearly outperform open-source ones. In terms of overall performance, GPT‑4o achieved the best results in English tasks (4.05), while Doubao ranked highest in Chinese tasks (4.10). By contrast, open-source models generally scored between 2 and 3 points. Among open-source systems, Kimi‑Audio (3.11) performed best in Chinese, while Step‑Audio (2.77) led in English. These results highlight that commercial models remain substantially stronger in speech generation. 
The performance gap arises mainly from two factors. From a technical perspective, open‑source models emphasize “content correctness” but lack robustness in expressive speech generation, as they mostly rely on semantic‑level representations and insufficiently model acoustic features. An exception is Baichuan‑Audio, which uses a unified codec to better capture vocal attributes, achieving strong results in age control and showing some timbre-control ability. From a resource perspective, commercial models benefit from larger training corpora and stronger computation, yielding greater stability, while open‑source systems often struggle with instruction-following, leading to frequent low scores that lower overall performance.

\textbf{Variation across task categories}. 
Models show clear differences across task types. In acoustic attributes, composite tasks requiring control of multiple dimensions are more difficult than single-attribute tasks and score lower overall. In instruction-following, GPT‑4o demonstrates strong English ability across all subtasks, including the hardest variation type, and shows robust style adaptation. Other models perform notably worse in style variation than in the other subtasks, indicating this remains difficult. In role-play, GPT‑4o performs well in scene and character subtasks, while Doubao is strong in Chinese scene tasks. In implicit empathy, several models handle emotions effectively, with both positive and negative emotions generally well processed.

\textbf{Language preference among models}. Performance varies significantly across languages. For example, Doubao, Kimi‑Audio, and Qwen2.5‑Omni perform much better in Chinese than in English, whereas the GPT‑4o series shows the opposite pattern. This discrepancy may stem from imbalanced distributions of training data across languages, or from substantial differences in pronunciation habits, which make cross-lingual transfer in speech generation far less effective than in text generation.

\subsection{Evaluation Consistency Analysis}
To assess the consistency between model scores and human evaluations, we randomly sampled approximately half of the instructions for human assessment. Prior to the formal study, all annotators were required to complete a trial task and pass a qualification test; only those who met the standard were eligible to participate. Each instance was independently rated by five expert annotators. To ensure consistency in evaluation criteria, annotators followed the same set of instructions and guidelines as those used in the model-based assessment.

\begin{table}[h]
\centering
\caption{Spearman Correlation Between Human and Model Evaluations (\%)}
\label{tab:correlation-analysis}
\begin{tabular}{lS[table-format=2.2]S[table-format=2.2]}
\toprule
\textbf{Metric} & \textbf{English (\%)} & \textbf{Chinese (\%)} \\
\midrule
Inter-Human               & 78.58 & 70.54 \\
Model-Individual Human    & 70.61 & 64.53 \\
Model-Consensus Human                   & 77.01 & 73.03 \\
\bottomrule
\end{tabular}
\end{table}

To measure consistency, we adopted Spearman's rank correlation coefficient.
Specifically:
\begin{itemize}
\item To assess the reliability of human evaluations, we calculated the average Spearman correlation coefficient across different annotators (\textbf{Inter-Human Agreement (Average)}).
\item To assess the alignment between the model and human judgments, we calculated the average Spearman correlation coefficient between the model and each individual annotator (\textbf{Model–Individual Human Agreement (Average)}), as well as the correlation between the model and the mean score of all annotators (\textbf{Model–Consensus Human Agreement}).
\end{itemize}

The results are presented in Table~\ref{tab:correlation-analysis}. The correlations between model scores and human judgments are generally close to inter-human agreement levels. Notably, the correlation with the average human score reached \textbf{77.01\%} for English and \textbf{73.03\%} for Chinese, comparable to the human agreement benchmark. These findings indicate that the model can reliably approximate human evaluation patterns with cross-lingual stability. Overall, model-based evaluation achieves near human-level consistency and can serve as an efficient and scalable alternative to manual assessment.

\section{Limitation and Conclusion}
While VStyle introduces a novel benchmark for voice style adaptation, several limitations remain. First, the instruction dataset is built from manually designed seeds and LLM‑based expansion, so its distribution reflects annotator preferences and model‑driven patterns, which may diverge from real user demands. Second, although Large Audio Language Models (LALMs) show strong potential in audio evaluation, they still face issues such as hallucinations. To address this, we adopt step‑by‑step prompts and explicit guidelines, which help improve consistency and reliability. Future advances in reasoning and auditory perception will further enhance the evaluation pipeline.

Despite these constraints, VStyle establishes an essential foundation for progress in voice style adaptation. On one hand, its bilingual and multi‑category design covers realistic interaction needs ranging from acoustic attribute control to implicit empathy, thus complementing existing benchmarks that often overlook expressive aspects. On the other hand, our experiments reveal strong correlations between LALM‑as‑a‑Judge and human assessment, confirming the scalability and practicality of our framework. We hope VStyle will serve not only as a diagnostic tool for identifying model shortcomings, but also as a catalyst for more natural, controllable, and human‑centered speech generation systems.

\label{sec:conclusion}

\bibliographystyle{IEEEbib}
\bibliography{strings,refs}

\end{document}